\documentclass[runningheads]{llncs}
\usepackage[T1]{fontenc}
\usepackage{amsmath}
\usepackage{graphicx}
\usepackage{adjustbox}
\usepackage{comment}
\usepackage{multirow}
\usepackage{xcolor}
\usepackage{hyperref}
\usepackage{color}

\begin{document}
\title{Generative Diffusion Model Bootstraps Zero-shot Classification of Fetal Ultrasound Images In Underrepresented African Populations}
\titlerunning{FU-LoRA: Generative Diffusion Model}
%
\author{Fangyijie Wang\inst{1,2}\orcidID{0009-0003-0427-368X} \and
Kevin Whelan\inst{2} \and
Gu\'enol\'e Silvestre\inst{1,3} \and
Kathleen M. Curran\inst{1,2}\orcidID{0000-0003-0095-9337}}
\authorrunning{F. Wang et al.}
%
\institute{
Science Foundation Ireland Centre for Research Training in Machine Learning 
\email{fangyijie.wang@ucdconnect.ie}\\ \and
School of Medicine, University College Dublin, Dublin, Ireland \and
School of Computer Science, University College Dublin, Dublin, Ireland
}
\maketitle              
\begin{abstract}

Developing robust deep learning models for fetal ultrasound image analysis requires comprehensive, high-quality datasets to effectively learn informative data representations within the domain. However, the scarcity of labelled ultrasound images poses substantial challenges, especially in low-resource settings. To tackle this challenge, we leverage synthetic data to enhance the generalizability of deep learning models. This study proposes a diffusion-based method, \textbf{F}etal \textbf{U}ltrasound \textbf{LoRA} (\textbf{FU-LoRA}), which involves fine-tuning latent diffusion models using the LoRA technique to generate synthetic fetal ultrasound images. These synthetic images are integrated into a hybrid dataset that combines real-world and synthetic images to improve the performance of zero-shot classifiers in low-resource settings. Our experimental results on fetal ultrasound images from African cohorts demonstrate that FU-LoRA outperforms the baseline method by a 13.73\% increase in zero-shot classification accuracy. Furthermore, FU-LoRA achieves the highest accuracy of 82.40\%, the highest F-score of 86.54\%, and the highest AUC of 89.78\%. It demonstrates that the FU-LoRA method is effective in the zero-shot classification of fetal ultrasound images in low-resource settings. Our code and data are publicly accessible on \href{https://github.com/13204942/FU-LoRA}{GitHub}.

\end{abstract}
\section{Introduction}

Ultrasound imaging is widely used for the diagnosis, screening and treatment of many diseases because of its portability, low cost and non-invasive nature \cite{Zaffino:2020}. In recent decades, ultrasound screening has become a common method used for prenatal evaluation of fetal growth, fetal anatomy, and estimation of gestational age (GA), as well as monitoring pregnancy \cite{Loughna:2009,Salomon:2011}. After 14 weeks of gestation, GA is estimated using standard measurements such as head circumference, biparietal diameter, occipito-frontal diameter, transcerebellar diameter, lateral ventricles, abdominal circumference, and femur length \cite{Fiorentino:2023}. These fetal biometric measurements are performed following a standardized procedure through the identification of the standard sonographic plane \cite{Salomon:2019}. An automated classification system can assist sonographers in promptly and accurately identify maternal–fetal standard planes. Recent deep learning methods have demonstrated significant potential for analyzing fetal ultrasound images. However, existing developments have mainly concentrated on applications in high-resource settings (HRS), where there is access to extensive clinical imaging datasets utilized for training deep learning models \cite{Piaggio:2021}. In low-resource settings (LRS), the scarcity of clinical images remains a significant challenge for the generalizability of deep learning models. LRS is characterized by a lack of adequate healthcare resources and systems that fail to meet recognized global standards \cite{Piaggio:2021}.

Utilization of diffusion models for synthetic image generation has the potential to enrich medical imaging datasets, especially in scenarios where data are limited in LRS and where increased diversity is essential in existing medical imaging modalities \cite{Kazerouni:2023}. The diffusion model has been used in various medical image enhancement applications, such as Computed Tomography (CT)~\cite{Gao:2024}, Positron Emission Tomography (PET) \cite{Jiang_pet:2023}, Magnetic resonance imaging (MRI)~\cite{Pinaya:2022}, X-ray~\cite{kim:2023}, and ultrasound~\cite{Gaona:2024}. Lee et al.~\cite{Lee:2020} propose an information maximizing generative adversarial network (GAN) to generate synthetic examples of fetal brain ultrasound images from twenty-week anatomy scans. Lasala et al.~\cite{Lasala:2024} introduce an approach that leverages class activation maps (CAM) as a prior condition to generate standard planes of the fetal head using a conditional GAN model. However, these studies only investigate the effectiveness of synthetic fetal head ultrasound images, and their methods are image-to-image based generation approaches. 

In this study, we aim to address the data scarcity of fetal ultrasound images in LRS. We propose a novel text-to-image generation approach that utilizes the diffusion model to enhance the performance of convolutional neural networks (CNN) for the classification of five common maternal fetal ultrasound planes. The proposed approach involves the following steps: i) Fine-tuning the diffusion model on a dataset collected in HRS to enable the pre-trained model to learn distinct features in fetal ultrasound data, ii) Creating hybrid datasets by integrating real-world data with synthetic data using the fine-tuned text-to-image diffusion model. 

Our contributions are: i) proposing a novel approach to improve zero-shot image classification accuracy in low-resource settings using synthetic data generated from latent diffusion models, ii) publishing the first LoRA model for the generation of synthetic fetal ultrasound images, and iii) releasing a synthetic dataset of 5000 images for further research in fetal ultrasound field.

\section{Methods}

\subsubsection{Preliminaries}

Latent Diffusion Model (LDM)~\cite{Rombach:2022}, stands out as one of the most successful diffusion models available within the current open-source community. Structurally, LDM is a Denoising Diffusion Probabilistic Model (DDPM) \cite{ho:2020} implemented in the latent space to efficiently decrease computational costs. 

Training the LDM involves a process that contains a diffusion (or forward) process and a sampling (or reverse) process. Given an image $x_0 \in X$, the diffusion process gradually adds random Gaussian noise to the input image $x_0$ at diffusion step $t$ ($T$ denotes the total number of diffusion steps) to produce $x_t$, following a Markov Chain by:
$
q\left(\boldsymbol{x}_t \mid \boldsymbol{x}_{t-1}\right)=\mathcal{N}\left(\boldsymbol{x}_t ; \sqrt{1-\beta_t} \boldsymbol{x}_{t-1}, \beta_t \boldsymbol{I}\right)
$
where $\beta_t \in [0, 1]$ represents the variance schedule across diffusion steps, $\boldsymbol{I}$ is the identity matrix, $x_t$ and $x_{t-1}$ are adjacent image status. Accordingly, a noisy target $x_t$ distribution from the data $x_0$ is represented as: $ x_t =\sqrt{\alpha_t} x_0+\sqrt{\left(1-\alpha_t\right)} \epsilon $
where $\alpha_t=\Pi_{s=1}^t\left(1-\beta_s\right)$. Then a U-Net \cite{Ronneberger:2015} is trained to approximate the reverse denoising process:
$
p_\theta\left(\boldsymbol{x}_{t-1} \mid \boldsymbol{x}_t\right)=\mathcal{N}\left(\boldsymbol{x}_{t-1} ; \boldsymbol{\mu}_\theta\left(\boldsymbol{x}_t, t\right), \beta_t \boldsymbol{I}\right)
$
where $\mu_\theta$ is a parameterized mean with the noise predictor $\epsilon_\theta$, which consists of a U-Net ($\theta$ denotes model parameters):

\begin{equation}
\boldsymbol{\mu}_\theta\left(\boldsymbol{x}_t, t\right)=\frac{1}{\sqrt{1-\beta_t}}\left(\boldsymbol{x}_t-\frac{\beta_t}{\sqrt{1-\alpha_t}} \boldsymbol{\epsilon}_\theta\left(\boldsymbol{x}_t, t\right)\right)
\end{equation}

\noindent The U-Net ($\epsilon_\theta$) is trained with the mean square loss 
$L := E_{t,x} ||\epsilon - \boldsymbol{\epsilon}_\theta(\boldsymbol{x}_t, t)||^2 $ 
where $\epsilon \sim N(0,\boldsymbol{I})$. In the sampling process, LDM learns the Markov chain to convert the Gaussian noise distribution $x_T \sim N (0,\boldsymbol{I})$ into the target distribution $x_0$ by the iterative denoising steps: $\boldsymbol{x}_{t-1}=\boldsymbol{\mu}_\theta\left(\boldsymbol{x}_t, t\right)+\beta_t \boldsymbol{z}$ where $\boldsymbol{z} \sim \mathcal{N}(0, \boldsymbol{I})$.
We follow Rombach's work \cite{Rombach:2022} to control the generation procedure by integrating the conditioning input, text $c$, to the noise predictor U-Net $\boldsymbol{\epsilon}_\theta(\boldsymbol{x}_t,t)$. Therefore, $\boldsymbol{\epsilon}_\theta$ is defined as $\boldsymbol{\epsilon}_\theta(\boldsymbol{x}_t,c,t)$ optimized according to the objective loss $L$ where $c$ represents the text prompts with CLIP (Contrastive Language-Image Pre-Training) encoding \cite{Radford:2021}.

The Low-Rank Adaption (LoRA) technique \cite{hu:2022} is designed for the efficient fine-tuning of large language models and can also be utilized for fine-tuning generative models. Therefore, we apply LoRA technique to the cross-attention layers within the U-Net architecture ($\boldsymbol{\epsilon}_\theta$ ) \cite{Ronneberger:2015}. This integration accelerates the training process, leading to decreased computational demands and a reduced model size. The formula of LoRA is defined as:
\begin{equation}
\label{rank_eq}
    h=W_0 x+\frac{\alpha}{r} \Delta W x=W_0 x+\frac{\alpha}{r} B A x
\end{equation}
where $W_0$ is a pre-trained weight matrix of U-Net within diffusion models. $W_0$ can be decomposed into two smaller matrices, denoted as $A$ and $B$, with a reduced rank $r$ compared to the original matrix. $\frac{\alpha}{r}$ represents the merging ratio and ranges from 0 to 1. 
During the training process, $W_0$ remains frozen, while the matrices $A$ and $B$ are equipped with trainable parameters. Matrix $A$ is initialized using Gaussian distribution, while matrix $B$ is initialized with zeros.

\subsubsection{FU-LoRA}
\label{lora_ft}

\begin{figure}[t]
    \centering
    \includegraphics[width=0.9\textwidth]{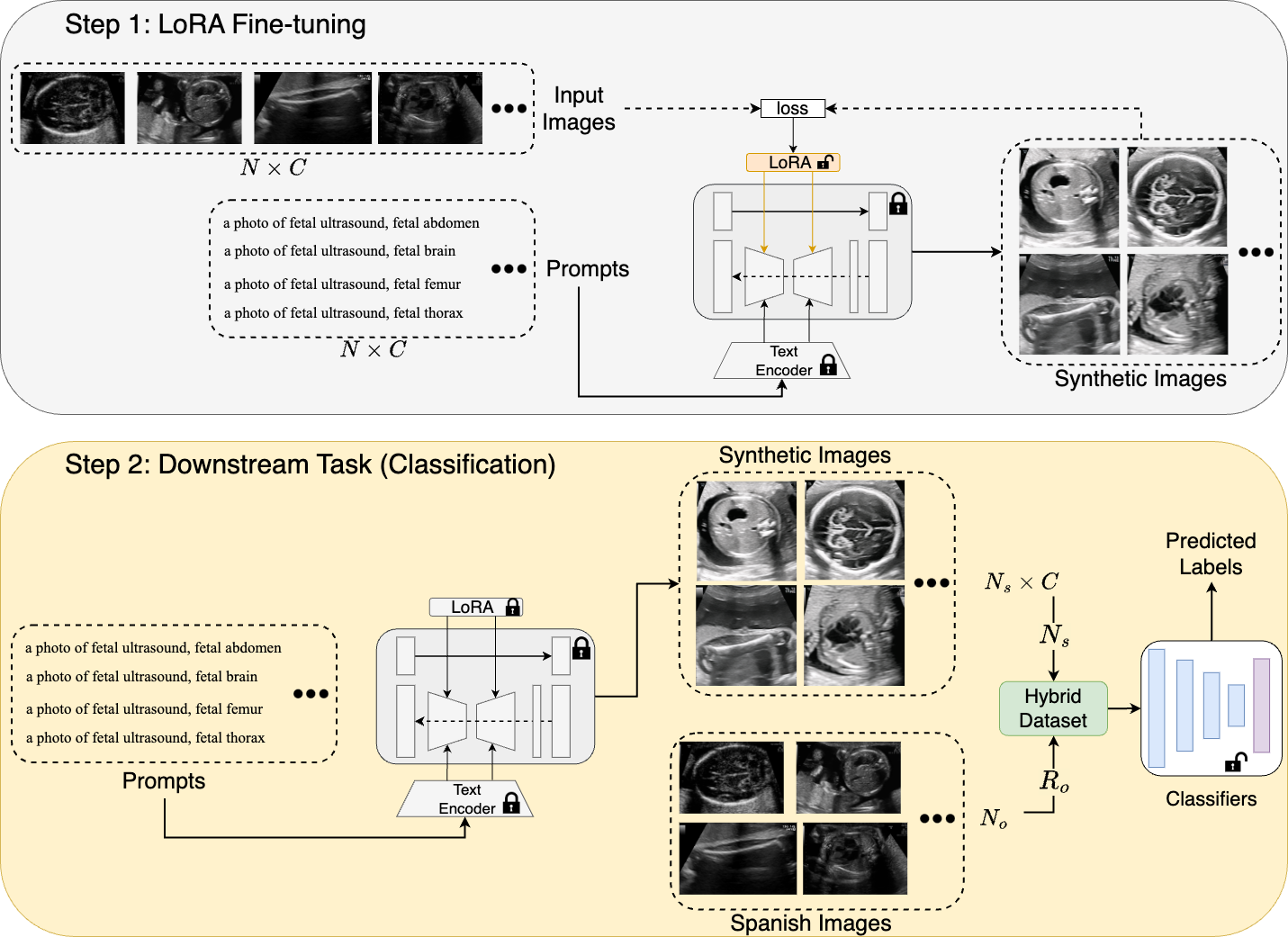}
    \caption{The overview of our proposed FU-LoRA method. $N$: number of Spanish data per plane for training LoRA; $C$: number of planes; $N_s$: number of synthetic images; $N_o$: total number of Spanish images; $R_o$: random subset of $N_o$.} 
    \label{fu_lora_approach}
\end{figure}

Fig. \ref{fu_lora_approach} shows the two steps involved in our FU-LoRA method: (1) Fine-tuning the pre-trained diffusion model using the LoRA method on a small fetal ultrasound dataset from HRS. (2) Employing the fine-tuned LoRA model for training downstream tasks in LRS. This approach integrates synthetic images to enhance generalization and performance of deep learning models. We conduct three fine-tuning sessions for the diffusion model to generate three LoRA models with different hyper-parameters: $\alpha \in [8, 32, 128]$, and $r \in [8, 32, 128]$. The merging rate $\frac{\alpha}{r}$ in Equation. \ref{rank_eq} is fixed to 1. The purpose of this operation is to delve deeper into LoRA to uncover optimizations that can improve the model's performance and evaluate the effectiveness of parameter $r$ in generating synthetic images. 

\subsubsection{Downstream Classification}

We aim to evaluate the quality and utility of the synthetic data generated by FU-LoRA. Therefore, we use them to train five classifiers, incorporating DenseNet169, ResNet18, EfficientNet b0, MobileNet v2, and Vision Transformer (ViT b16), which are subsequently tested on African data. The final fully connected layers of these models are substituted with new fully connected layers having an output dimension of 5. Initially, all models are pre-trained on ImageNet~\cite{Russakovsky:2015} and subsequently fully re-trained throughout the entire network.

\subsubsection{Evaluation Metrics}

The generation performance of LoRA with different values of rank $r$ in Equation. \ref{rank_eq} is assessed by evaluating the diversity of the generated samples using t-distributed stochastic neighbor embedding (t-SNE) \cite{Maaten:2008}. t-SNE is a statistical technique used for visualizing high-dimensional data, assigning each data point a specific location on a two or three-dimensional map. We extract ultrasound image features using the InceptionNet \cite{Szegedy:2016} pre-trained on ImageNet and visualize them by t-SNE in a two-dimensional map (see Section \ref{syn_image}). In the African dataset, the fetal thorax plane contains 40 images. To ensure fair evaluation, we randomly select 35 images from the Spanish, Synthesis, and African datasets for four anatomical planes: Abdomen, Brain, Femur, and Thorax. Thus, a total of 140 images are gathered for qualitative analysis. 

\section{Materials and Experiments}

\subsubsection{Datasets:}

We utilize two public datasets and one synthetic dataset in this study: i) the Spanish dataset is acquired from two centers in Spain \cite{Xavier:2020}, ii) the African dataset is acquired from Algeria, Egypt, Malawi, Ghana, and Uganda \cite{sendra_balcells:2023}, and iii) the synthetic fetal ultrasound dataset is generated by the FU-LoRA method. The variability of the fetal head across various pregnancy trimesters poses a challenge in developing a robust deep learning model.

The Spanish dataset\footnote{https://zenodo.org/records/3904280} in HRS includes 1,792 patient records in Spain \cite{Xavier:2020}. All images are acquired during screening in the second and third trimesters of pregnancy using six different machines operated by operators with similar expertise. We randomly selected 20 Spanish ultrasound images from each of the five maternal–fetal planes (Abdomen, Brain, Femur, Thorax, and Other) to fine-tune the LDM using LoRA technique, and 1150 Spanish images (230 $\times$ 5 planes) to create the hybrid dataset. In summary, fine-tuning the LDM utilizes 100 images including 85 patients. Training downstream classifiers uses 6148 images from 612 patients. Within the 6148 images used for training, a subset of 200 images is randomly selected for validation purposes. The hybrid dataset employed in this study has a total of 1150 Spanish images, representing 486 patients.

The African dataset\footnote{https://zenodo.org/records/7540448} in LRS contains 450 images (125 patients) from five African countries \cite{sendra_balcells:2023} collected with five different ultrasound machines during the second and third pregnancy trimesters.This dataset only contains four maternal–fetal standard planes, including Abdomen, Brain, Femur, and Thorax. A total of 217 images from 61 patients can be used for training, while 233 images from 66 patients are allocated for testing purposes. To evaluate generalization of classification models with synthetic data, the 217 training images are only used for ablation studies as detailed in Section \ref{ab_study}. The 233 testing images, on the other hand, are utilized for evaluating the performance of downstream classifiers, as discussed in Section \ref{plane_clf}.

We create the synthetic dataset comprising 5000 fetal ultrasound images (500 $\times$ 2 samplers $\times$ 5 planes) accessible to the open-source community. The generation process utilizes our LoRA model Rank $r = 128$ with Euler \cite{karras:2022} and UniPC \cite{liu:2024} samplers known for their efficiency. Subsequently, we integrate this synthetic dataset with a small amount of Spanish data to create a hybrid dataset, shown in Fig. \ref{fu_lora_approach}.

\subsubsection{Traditional Data Augmentation:} 

The traditional data augmentation techniques are employed to train downstream classifiers for performance comparison with our methods. These techniques are: rotation by an angle from $[-25^{\circ}, 25^{\circ}]$, horizontal flipping with 50\% probability ($P(\cdot) = 0.5$), vertical flipping with 10\% probability ($P(\cdot) = 0.1$), and pixel normalization to float precision in $[-1, 1]$ range. All images are resized to 512 $\times$ 512 pixels for training, validation and testing purposes.

\subsubsection{Implementation Details:} 

The hyper-parameters of LoRA models are defined as follows:  batch size to 2; training epochs to 1; LoRA learning rate to $1e-4$; total training steps to 10000 ($100$ images $\times$ $100$ steps $\times$ $1$ epoch); LoRA dimension to 128; mixed precision selection to fp16; learning scheduler to constant; and input size (resolution) to 512. The model is trained on a single NVIDIA RTX A5000, 24 GB with 8-bit Adam optimizer on PyTorch. The training hyper-parameters of downstream classification models are: epochs to 20, batch size to 24, and learning rate to $1e-3$. The loss function computes the cross-entropy loss between input logits and target. All training processes are conducted on a single NVIDIA RTX 4090, 24 GB with stochastic gradient descent (SGD) optimizer on PyTorch. The input pixels of all images are converted from integer in the range $[0, 255]$ to single float precision in $[-1, 1]$ for training and testing the downstream classification models.

\begin{figure}[ht]
    \centering
    \includegraphics[width=0.8\textwidth]{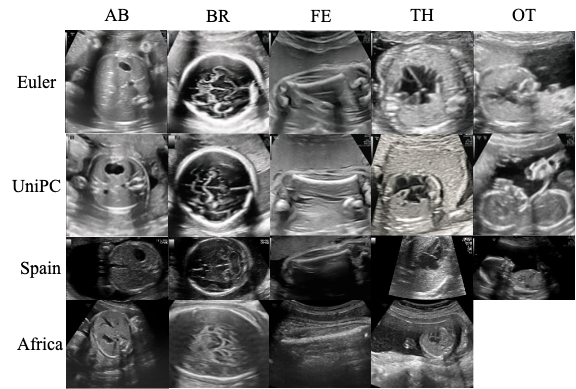}
    \caption{Examples of synthetic images generated using the FU-LoRA method are presented. The first two rows display images generated by the Euler sampler and the UniPC sampler, while the third and fourth rows show Spanish and African images utilized for fine-tuning the diffusion model. AB: Abdomen; BR: Brain; FE: Femur; TH: Thorax; OT: Other.} 
    \label{sample_outputs}
\end{figure}

\begin{figure}[!h]
    \centering
    \includegraphics[width=\textwidth]{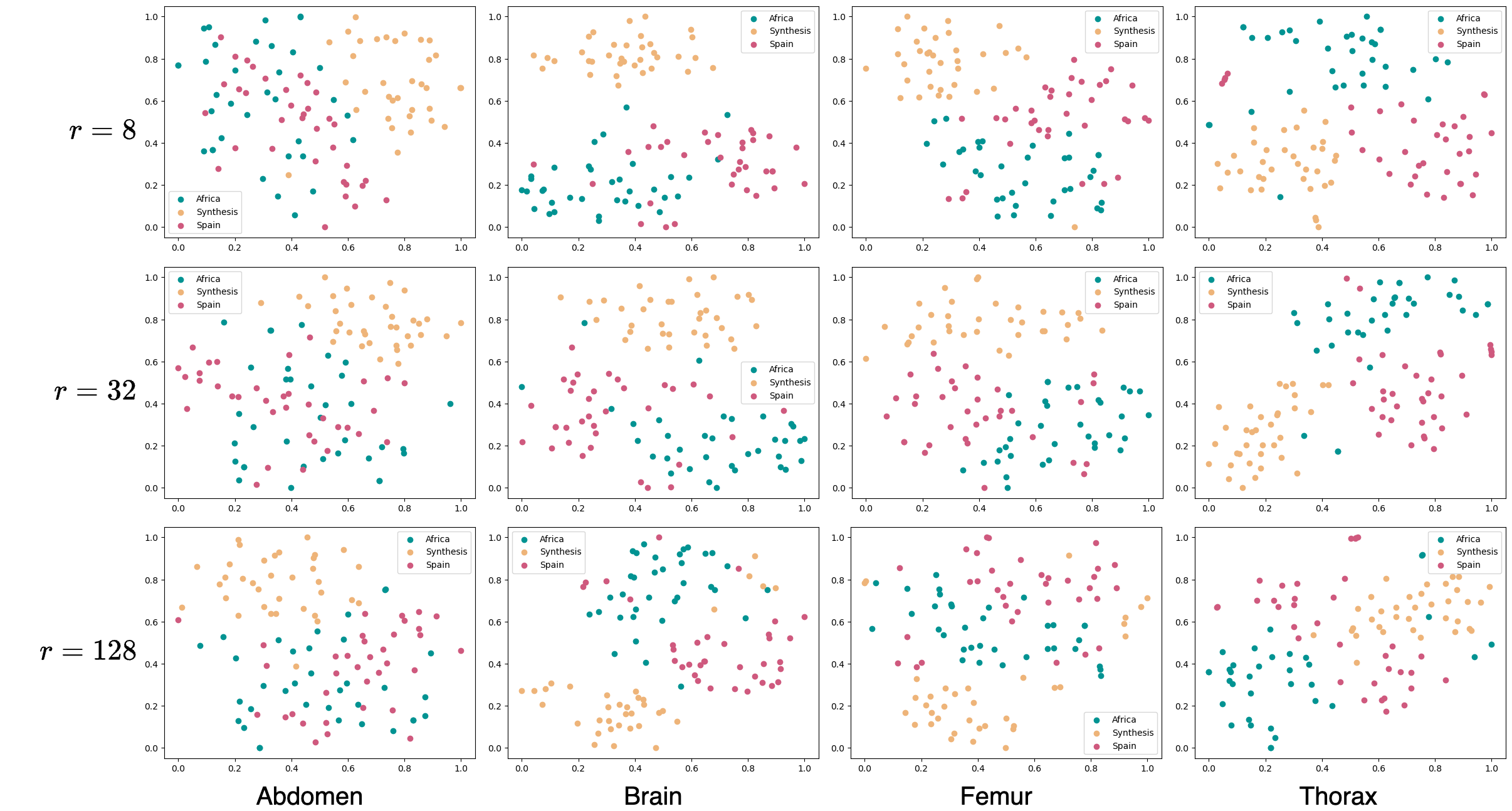}
    \caption{An overview of the feature embeddings of synthetic images generated by LoRA models (rank $r \in [128,32,8]$) using Spanish and African images through t-SNE visualization. The feature embeddings are extracted using a pre-trained InceptionNet model. From left to right, the t-SNE visualization is applied to Abdomen, Brain, Femur, and Thorax plane. Green: Africa; Yellow: Synthesis; Red: Spain.}
    \label{tsne_vis}
\end{figure}

\section{Results}

\subsubsection{Synthetic Images:}
\label{syn_image}

The process of generating synthetic images from text is commonly known as text-to-image generation. We utilize two sampling methods: Euler \cite{karras:2022} and UniPC \cite{liu:2024}. During the process, the weights of LoRA are fixed at $1.0$, and the number of sampling steps is maintained at 20. As a result, we generate a hybrid dataset of 5000 images by utilizing the LoRA Rank 128 model to generate 2500 synthetic images using Euler and UniPC samplers, respectively.

Fig. \ref{sample_outputs} presents the samples of the synthetic fetal ultrasound images generated with our proposed FU-LoRA method in Fig. \ref{fu_lora_approach} by providing text prompt inputs for each maternal plane. It shows that the fetal abdominal images depict the stomach situated within the abdomen. The skull is prominently visible in the brain images as a bright, elliptical structure. Consequently, the femur images illustrate a single fetal femur with high brightness. The fetal thorax images feature the right atrium, left atrium, and both ventricles. 

The t-SNE visualization in Fig. \ref{tsne_vis} offers a qualitative assessment of the overall quality of the generated images. Within this two-dimensional embedded space, the real (green and red) and synthetic (yellow) data are positioned in distinct regions in Abdomen, Brain and Femur planes with rank $r=8$. There are obvious separations in Brain and Thorax planes with rank $r=32$. The lack of separation in the feature space in Abdomen, Femur and Thorax with rank $r=128$. Compared with the other two LoRA models with lower rank $r$ as shown in the second and third rows of Fig. \ref{tsne_vis}, the model with rank $r=128$ generates data points that are located closer to those regions of real-world data.

\begin{table}[!t]
    \centering
    \caption{An overview of the classification performance of multiple deep learning models on the African fetal ultrasound dataset. ES: Spanish dataset; HB: Hybrid dataset (Spanish and Synthetic images); DA: Traditional data augmentation techniques.}    
    \begin{adjustbox}{width=\linewidth}
    \begin{tabular}{llcccccccc}
         \hline
         \multirow{2}{*}{Data} & \multirow{2}{*}{Model} & \multicolumn{3}{c}{Training Data} & \multirow{2}{*}{ACC} & \multirow{2}{*}{Recall} & \multirow{2}{*}{Precision}  & \multirow{2}{*}{F-score} & \multirow{2}{*}{AUC} \\
         & & \# ES & \# Synthetic & \# Patients & & & & \\
         \hline
         ES & DenseNet169 & 6148 & 0 & 612 & 61.37 & 62.38 & 96.46 & 73.31 & 80.50 \\         
         ES & MobileNet v2 & 6148 & 0 & 612 & 68.67 & 68.86 & 96.50 & 78.07 & 83.99 \\
         ES & EfficientNet b0 & 6148 & 0 & 612 & 63.09 & 62.64 & 96.94 & 74.34 & 81.15 \\
         ES & ResNet18 & 6148 & 0 & 612 & 54.94 & 51.23 & 90.95 & 61.51 & 76.09 \\
         ES & ViT b16 & 6148 & 0 & 612 & 68.24 & 67.05 & 96.58 & 77.53 & 83.46 \\         
         \hline
         ES+DA & DenseNet169 \cite{sendra_balcells:2023} & 6148 & 0 & 612 & 68.67 & 67.67 & 92.43 & 75.87 & 82.52 \\
         ES+DA & MobileNet v2 & 6148 & 0 & 612 & 69.96 & 69.05 & 95.27 & 78.95 & 83.93 \\
         ES+DA & EfficientNet b0 & 6148 & 0 & 612 & 59.66 & 59.26 & 94.72 & 70.91 & 79.09 \\
         ES+DA & ResNet18 & 6148 & 0 & 612 & 63.09 & 61.10 & 91.52 & 71.01 & 79.86 \\
         ES+DA & ViT b16 & 6148 & 0 & 612 & 74.68 & 72.80 & 97.05 & 82.15 & 86.23 \\
         \hline         
         HB & DenseNet169 & 1150 & 5000 & 486 & 75.54 & 73.40 & \textbf{98.66} & 83.40 & 77.04 \\
         HB & MobileNet v2 & 1150 & 5000 & 486 & 67.81 & 65.18 & 96.52 & 76.63 & 75.78 \\
         HB & EfficientNet b0 & 1150 & 5000 & 486 & 72.96 & 69.81 & 97.50 & 79.45 & 77.65 \\
         HB & ResNet18 & 1150 & 5000 & 486 & 64.81 & 60.15 & 96.92 & 69.95 & 76.88 \\
         HB & ViT b16 & 1150 & 5000 & 486 & 72.10 & 70.28 & 96.81 & 80.57 & 76.26 \\         
         \hline
         HB+DA & DenseNet169 & 1150 & 5000 & 486 & 80.26 & 77.61 & 89.11 & 82.59 & 87.99 \\
         HB+DA & MobileNet v2 & 1150 & 5000 & 486 & 77.68 & 74.29 & 86.69 & 79.36 & 86.11 \\
         HB+DA & EfficientNet b0 & 1150 & 5000 & 486 & 72.96 & 68.75 & 92.65 & 75.91 & 83.87 \\
         HB+DA & ResNet18 & 1150 & 5000 & 486 & \textbf{82.40} & \textbf{81.56} & 89.15 & 84.90 & 89.29 \\
         HB+DA & ViT b16 & 1150 & 5000 & 486 & 81.97 & 79.92 & 95.36 & \textbf{86.54} & \textbf{89.78} \\
         \hline
    \end{tabular}
    \end{adjustbox}
    \label{tab:test_diff_ft_methods}
\end{table}

\begin{table}[!t]
    \centering
    \caption{Effectiveness of utilizing various data models to train ViT b16 to identify maternal-fetal standard planes in African populations. ES: Spanish dataset; AF: African dataset; HB: Hybrid dataset. ACC: Accuracy. AUC: Area under the ROC Curve.}    
    \begin{adjustbox}{width=\linewidth}
    \begin{tabular}{lc|cccc|ccccc}
         \hline
         \multirow{2}{*}{Data} & \multirow{2}{*}{Model} & \multicolumn{4}{c|}{Training Data} & \multirow{2}{*}{ACC} & \multirow{2}{*}{Recall} & \multirow{2}{*}{Precision} & \multirow{2}{*}{F-score} & \multirow{2}{*}{AUC} \\
         & & \# ES & \# AF & \# Synthetic & \# Patients & & & & \\
         \hline
         AF & ViT b16 & 0 & 217 & 0 & 61 & 91.41 & 88.71 & 93.23 & 90.44 & 93.81 \\
         ES & ViT b16 & 6148 & 0 & 0 & 612 & 68.24 & 67.05 & 96.58 & 77.53 & 83.46 \\
         ES+AF & ViT b16 & 6148 & 217 & 0 & 673 & \textbf{92.27} & \textbf{91.53} & 96.67 & \textbf{93.87} & \textbf{95.40} \\
         \hline
         HB & ViT b16 & 6148 & 0 & 5000 & 612 & 78.54 & 78.14 & \textbf{97.44} & 85.85 & 88.76 \\
         \hline
    \end{tabular}
    \end{adjustbox}
    \label{tab:ablation_res}
\end{table}

\subsubsection{Effectiveness Evaluation:}
\label{plane_clf}

We evaluate the model's classification performance using synthetic data to classify fetal anatomical planes. Notably, African data are used only to test classification models. The quantitative results of five classification models are provided in Table \ref{tab:test_diff_ft_methods}. It is observed that utilizing the $HB+DA$ data model yields highly comparable accuracy for anatomical plane classifiers. Among the five classifiers, ViT achieves the highest F-score of 86.54\% and AUC of 89.78\% on African data. More importantly, we observe that $HB+DA$ [ViT] achieves a significant improvement in AUC of 7.2\% compared to the model $ES+DA$ [DenseNet169] used by other researchers \cite{sendra_balcells:2023}. The average F-score across all models trained with $HB+DA$ is 81.86\%, whereas $HB$, $ES$ and $ES+DA$ data models have average F-score of 78\%, 72.95\% and 75.78\%, respectively. Compared to the $ES$ [MobileNet] and $ES+DA$ [ViT] data models, the $HB$ [DenseNet169] increases the best classification accuracy by 6.87\% and 0.86\%. Notably, the $HB+DA$ [ResNet18] has 13.73\% higher accuracy than the baseline model $ES+DA$ [DenseNet169] \cite{sendra_balcells:2023}.

\subsubsection{Ablation Studies:}
\label{ab_study}

To verify the effectiveness of our proposed method, FU-LoRA, we compare our hybrid data model with the other three data models. Because the ViT b16 model achieves the highest F-score and AUC in Table \ref{tab:test_diff_ft_methods}, it is selected to train and test with the same experimental settings for ablation studies. Their quantitative results are given in Table \ref{tab:ablation_res}. Our observations are: (1) Our $HB$ data model has better zero-shot classification performance than the $ES$ data model across all metrics using 1150 Spanish images, a significantly smaller quantity than the original training set. (2) Compared to $AF$ data model, $HB$ exhibits a reduction in F-score of 4.6\% and AUC of 5\%, respectively, in the absence of target domain data during training. (3) Compared to the best case scenario of having $ES+AF$ for training, our $HB$ data model has acceptable differences across all evaluation metrics. Our key finding reveals that in the context of zero-shot classification, our $HB$ data model exhibits superior performance and generalization compared to the $ES$ data model when African data is not included for training classifiers.

\section{Limitations}

Our proposed method showcases promising results in generating realistic and anatomically meaningful synthetic images based on textual information. Nevertheless, our current work still has limitations. This study integrates text prompts to generate accurate and medically significant images. Integrating more detailed texts could improve the diffusion model. Moreover, Figure \ref{tsne_vis} shows the separation in the feature embeddings of the Brain and Femur planes with rank $r=128$. It suggests that synthetic images may lack semantically relevant features. Higher-quality synthetic images could be generated using alternative fine-tuning methods such as DreamBooth \cite{Ruiz:2023} and Textual Inversion \cite{gal:2022}. However, our study focuses on the feasibility of the most efficient fine-tuning method, LoRA, which involves the fewest images. We aim to address these limitations in future work.

\section{Conclusion}

In this study, we have effectively fine-tune the LDM to produce synthetic fetal ultrasound images that accurately reproduce the characteristics observed in real-world images. Our proposed method, FU-LoRA, requires only 100 fetal ultrasound images from 85 patients for the effective process of fine-tuning. Additionally, we demonstrate that FU-LoRA method can facilitate zero-shot classification for standard fetal ultrasound planes in obstetric ultrasound within low-resource settings. Using a hybrid dataset comprising synthetic images generated by FU-LoRA for training deep learning models, we achieve the highest F-score of 86.54\% and the highest AUC of 89.78\% in classifying African standard fetal planes. We publicly release a synthetic dataset to address a significant challenge in ultrasound image analysis: the scarcity of extensive annotated fetal ultrasound data, all while safeguarding privacy. In conclusion, our results highlight the potential of the FU-LoRA method for zero-shot learning in fetal ultrasound imaging analysis within low-resource settings.

\subsubsection{Acknowledgments.} This publication has emanated from research conducted with the financial support of Science Foundation Ireland under Grant number 18/CRT/6183. For the purpose of Open Access, the author has applied a CC BY public copyright licence to any Author Accepted Manuscript version arising from this submission.


%
%
%
\bibliographystyle{splncs04}
\bibliography{mybibliography}
%
%
%
%
%
\end{document}